\documentclass[preprint,showpacs,amssymb,aps,pra]{revtex4}

\begin{document}

\title{Effective Field Theory 
Program \\
for Conformal Quantum Anomalies}

\author{Horacio E. Camblong,$^{1}$
Luis N. Epele,$^{2}$
Huner Fanchiotti,$^{2}$ 
Carlos A. Garc\'{\i}a Canal,$^{2}$ and
Carlos R. Ord\'{o}\~{n}ez$^{3}$}

\affiliation{
$^{1}$
Department of Physics, University of San Francisco, San
Francisco, California 94117-1080, USA
\\
$^{2}$ 
Laboratorio de F\'{\i}sica Te\'{o}rica,
 Departamento de F\'{\i}sica,
Facultad de Ciencias Exactas, Universidad Nacional de La Plata,
 C.C. 67 -- 1900 La Plata, Argentina
\\
$^{3}$ Department of Physics, University of Houston, Houston,
Texas 77204-5506, USA
}

\begin{abstract}
The emergence of conformal states is established for any problem 
involving a domain of scales where the long-range SO(2,1)
conformally invariant interaction is applicable. 
Whenever a clear-cut separation of ultraviolet
and infrared cutoffs is in place, this renormalization mechanism
is capable of producing binding in the strong-coupling regime.
A realization of this phenomenon, in the form 
of dipole-bound anions, is discussed.
\end{abstract}


\maketitle

\section{Introduction}
\label{sec:introduction}

The renormalization program~\cite{weinberg}
provides an insightful framework for 
the description of physical scales within a given problem.
This assumes that the
characteristic dimensional scales are sufficiently separated,
as required by {\em effective field theory\/}~\cite{weinberg,polchinski}.
Moreover, symmetry considerations usually
furnish further analytical control over what
contributing factors might be relevant for the hierarchy of scales.
In addition to the well-known examples 
in high-energy physics and condensed matter physics,
an effective renormalization of a system in molecular physics
was introduced in Ref.~\cite{molecular_dipole_anomaly}, where
a symmetry-centered approach was developed for
the formation of dipole-bound anions by electron capture.
In the relevant domain of scales, the dominant physics---governed 
by an inverse square potential~\cite{camblong:isp_letter}---takes 
a scale-invariant form known as {\em conformal quantum mechanics\/}.

The central purpose of this paper is to develop an effective field 
theory program for the quantum anomaly
of Ref.~\cite{molecular_dipole_anomaly}.
Specifically, we address the role played by additional degrees of
freedom---for example, the rotational ones in the molecular case. 
In this manner,
we extensively use recent work on the renormalization and anomalous symmetry
breaking of conformal quantum mechanics~\cite{camblong_CQM}.
As a consequence, we will establish the following
results.

(i)
The conformal analysis is robust and fairly insensitive to the 
ultraviolet and infrared physics.

(ii) The effective field approach---centered on renormalization 
techniques---sheds light, e.g.,
on the properties of dipole-bound anions; this is 
in sharp contrast with the statements of Ref.~\cite{bawin}.

(iii)
The origin of a critical dipole moment for 
binding can be directly traced to the 
conformal interaction. 

In short, the predictions of 
the conformal framework of
Ref.~\cite{molecular_dipole_anomaly} are 
{\em not significantly altered by the inclusion of
additional degrees of freedom.\/}
Moreover,
a similar analysis can be applied to other problems 
for which the conformal quantum anomaly is relevant, for example,
for the Efimov effect~\cite{efimov_effect}.

\section{Conformal Quantum Mechanics and Dipole-Bound States}
\label{sec:CQM_dipole-bound}

In this section, we start by summarizing the results of
Ref.~\cite{molecular_dipole_anomaly}
for dipole-bound states in the language of
effective field theory~\cite{camblong_CQM}.
As we will see in the next section, the effective field
approach also provides the natural connection between this work
and the standard results of rotationally adiabatic
theory~\cite{Garrett_dip_rot,Clary_dip_rot,desfrancois_epj:98}.

\subsection{Conformal Physics of Dipole-Bound States}
\label{sec:conformal_dipole-bound}

The dominant part of the electron-molecule interaction 
can be described with a point dipole---the electron does
not significantly probe radial scales smaller than the 
size $a$ of the molecule.
Then, in three spatial dimensions,
the associated anisotropic Hamiltonian reads
\begin{equation}
H 
= 
\frac{ p^{2}}{2 \, m_{e} }  
- \frac{g}{r^{2}} \, \cos \theta
\; ,
\label{eq:ISP_Hamiltonian_unregularized_anisotropic}
\end{equation}
in which the coupling $g$ can be recast into a dimensionless
form $ \lambda = 2 m_{e} \, g/\hbar^{2}$.
Under time reparametrizations, 
this system displays an
SO(2,1) conformal symmetry,
whose breaking at the quantum-mechanical level
can be interpreted as a {\em quantum anomaly\/}.
As a first step, introducing separation of variables:
$\Psi (r,\Omega)
= 
u(r) \, \Xi (\Omega)/r$ in spherical coordinates.
This leads to a scale-invariant radial equation
\begin{equation}
\frac{d^{2}  u(r)}{dr^{2}} + 
\biggl[
 k^{2} 
+ \frac{\gamma (\lambda) }{r^{2}}
\biggr]
 u(r) = 0 
\; 
\label{eq:radial_eq_anisotropic}
\end{equation}
coupled to 
a scale-independent angular operator equation
\begin{equation}
\hat{A} (\lambda)
\,
\Xi (\Omega)
=
\gamma (\lambda)
\,
\Xi (\Omega)
 \; ,
\label{eq:angular_eq_anisotropic}
\end{equation}
where the eigenvalue  $\gamma \equiv \gamma (\lambda) $
plays the role of a  separation constant and
\begin{equation}
\hat{A} (\lambda)
= - {\bf l}^{2} +
\lambda
\,
\cos \theta
 \; ,
\label{eq:conformal_angular_operator}
\end{equation}
with ${\bf l}$ being the relative orbital angular momentum
of the electron about the molecule.
The problem defined 
by the equations above
is completely characterized by the 
solutions of conformal quantum mechanics.

\subsection{Radial Conformal Quantum Mechanics}
\label{sec:radial-problem}

Conformal quantum mechanics applies to the description of the radial 
problem.
All the properties and conclusions discussed herein 
rely on the existence of a domain of scales in which the 
dominant physics is scale invariant. 
A symmetry-centered analysis 
in the relevant conformally invariant domain
shows that the theory retains the SO(2,1) symmetry
at the quantum level when $\gamma < 1/4$,
with  $\gamma = 1/4$ being a critical point of the conformal framework.
The existence of a {\em conformal critical point\/}
\begin{equation}
\gamma^{(*)} 
\equiv
\gamma 
\bigl(
\lambda^{(*)} 
\bigr) 
= 1/4
\label{eq:conformal_critical_point}
\end{equation}
is the crucial ingredient that explains the experimental 
fact that electron binding by molecular anions
only occurs for dipole moments greater
than a critical value $p^{(*)}$~\cite{molecular_dipole_anomaly}.

Conformal quantum mechanics
is singular for  $\gamma \geq 1/4$, but can be rescued by the
use of renormalization, which yields 
 {\em conformal bound states\/}  with energies
$
E_{ n }
=
E_{0}
\,
\exp 
\left( - 2 \pi n/\Theta  \right)
$, 
 where $n$ is a positive integer,
$E_{0}$ is the arbitrary ground-state energy, and the conformal
parameter $\Theta$ is derived from the coupling according
to the rule~\cite{camblong_CQM}
\begin{equation}
\Theta = \sqrt{\gamma - \frac{1}{4}}
\; .
\label{eq:Theta_definition}
\end{equation}
The specific value of the characteristic scale $E_{0}$ defined by
the conformal tower of states
is sensitive to the ultraviolet physics and cannot
be predicted by a renormalization approach alone.
However, the scale $E_{0}$
is not relevant in the determination of the
relative values of bound-state energies,
as exhibited by the {\em geometric scaling\/} 
\begin{equation}
\frac{ E_{n'} }{ E_{n} } 
=  
\exp 
\left[
 - \frac{2 \pi  (n'-n )}{\Theta } 
\right]
\; ,
\label{eq:ratios_cutoff_BS_regularized_energies_phenomenological}
\end{equation}
which is a remnant of the original scale invariance.
In particular, the geometric ratio $e^{-2\pi/\Theta}$
of adjacent energy levels has a universal form that is {\em independent 
of the cutoff and impervious to the ultraviolet physics\/}. 
Finally,
the conformal states are characterized by 
normalized radial wave functions of the form
\begin{equation}
u (r)
=
\kappa
\,
\sqrt{ \frac{ 2 \sinh \left( \pi \Theta \right) }{ \pi \Theta } }
\,
\sqrt{r}
\;
K_{i\Theta} (\kappa r)
\; ,
\label{eq:radial_wave_function}
\end{equation}
where $K_{i\Theta} (z)$ is the Macdonald function of imaginary
index~\cite{macdonald_function}.
This is the function whose properties guarantee the
universal geometric
scaling~(\ref{eq:ratios_cutoff_BS_regularized_energies_phenomenological}).
In addition, the same function leads to an estimate of the characteristic
radial size of the electron probability distribution,
given by $\kappa^{-1}$, with relative ratios
$\kappa_{n}/\kappa_{n'} =  e^{   \pi  (n'-n )/\Theta }$
exhibiting a similar kind of universal geometric scaling.

In short, the generic properties of conformal
quantum mechanics determine the nature of the bound
states of molecular anions and are parametrized by the possible values of
the conformal parameter $\Theta$. In turn,
$\Theta$ is described, from Eq.~(\ref{eq:Theta_definition}),
in terms of the effective coupling
 $\gamma
=
\Theta^{2} + 1/4 $,
 which is completely determined by the
angular dependence of the interaction, through
 the eigenvalue equation~(\ref{eq:angular_eq_anisotropic}).
This is the problem to which we now turn.

\subsection{Angular Eigenvalue Equation}
\label{sec:angular-problem}

The angular problem for an anisotropic conformal interaction
is given by
Eq.~(\ref{eq:angular_eq_anisotropic}),
whose secular-determinant form
$
D (\gamma, \lambda) 
\equiv 
\det M (\gamma,\lambda) 
=
0
$
involves the infinite matrix
$
M(\gamma, \lambda) 
= - A(\lambda) + \gamma  \, \openone 
$,
with $\openone$ being the identity matrix.
In particular, in the angular momentum basis 
$\left| l,m \right\rangle$, 
the matrix elements
$\left\langle
l { m}
|
M (\gamma, \lambda)
|
l' { m}'
\right\rangle
=
 \delta_{mm'}
M_{ll'} (\gamma, \lambda; m)
$
 are diagonal with respect to $m$,
with tridiagonal blocks
\begin{equation} 
M_{ll'} (\gamma, \lambda; m)
=
\biggl[
l(l+1) 
+
\gamma
\biggr] 
\,  \delta_{ll'} 
-
\lambda 
\biggl[
{ N}_{l}(m)
\, \delta_{l, l'-1} 
+
\left( l \leftrightarrow l' \right)
\biggr]
\; ,
\label{eq:conformal_tridiagonal_Mll'}
\end{equation}
where
$
{ N}_{l}(m)
 = \sqrt{ [(l+1)^{2} -m^{2}]/[(2l+1)(2l+3)] }
$.
As a result,
the secular determinant
takes the factorized form
$D(\gamma, \lambda) = \Pi_{m} D_{m}(\gamma, \lambda) $
and the eigenvalues are given by 
the roots of the reduced determinants
$
D_{m}(\gamma, \lambda) \equiv
 \det 
\bigl[
M_{ll'} (\gamma, \lambda; m) 
\bigr]
= 0
$,
for all integer values of $m$.
At this purely conformal level, 
for every $m$,
the roots $\gamma_{h,m}$ 
can be arranged in a decreasing sequence: 
$
\gamma_{0,m} \geq  \gamma_{1,m}
\geq \gamma_{2,m} \geq \ldots
$,
with $h =0,1, \dots$,
and compared against the condition
for conformal criticality:
 $\gamma = \gamma^{(*)}=1/4$.
Equation~(\ref{eq:conformal_tridiagonal_Mll'})
implies the following trends:
  $\gamma$ is a monotonic
function with respect to both $\lambda$ and $m$,
increasing with $\lambda$ and decreasing with $m$.
In particular, for any finite dipole moment
$p$ (i.e., finite $\lambda$),
there exist only a finite number of {\em supercritical\/}
values of $\gamma$; in turn, for each $\gamma$,
there is an infinite {\em tower of conformal states\/}---possibly  
limited by the onset of nonconformal physics
for long-distance scales.
Hence the conformal bound states are completely 
characterized by the set of quantum numbers $(n,h,m)$, in which 
the subset $(h,m)$ determines $\gamma_{h,m}$, while  
$n$ labels the ordering of the conformal tower or geometric 
scaling.
The existence of these states in the ``supercritical regime'' 
yields anomalous breaking of the SO(2,1)
commutator algebra~\cite{camblong_CQM}.

An important related question is:
for the largest root $\gamma_{0,0}$,
what is the value $\lambda^{(*)}$ that generates a conformal
critical point?
By setting
$\gamma_{0,0} =
\gamma^{(*)} = 1/4$,
the ``principal conformal critical coupling''
becomes
$
\lambda^{(*)}_{\rm conf} \approx 1.279
$
whence the required critical dipole moment is
$p^{(*)}= p_{0} \, \lambda^{(*)} 
\approx 1.625$ D~\cite{fer:47,lev:67,molecular_dipole_anomaly}.
Likewise, for each of the other roots
$\gamma_{h,m}$,
the criticality condition
$\gamma_{h,m} =
\gamma^{(*)} = 1/4$
defines additional, increasingly larger  values 
$\lambda ^{(*)}_{h,m}$ 
of the critical dipole moment.
Each of these represents the onset of a new tower
of conformal states of the
form~(\ref{eq:ratios_cutoff_BS_regularized_energies_phenomenological}).
The sequence of critical values of the dipole moment includes
$\lambda_{0,0}^{(*)} \approx 1.279;
\lambda ^{(*)}_{0,1} \approx  7.58; \ldots $.
However, the experimentally observed bound 
states~\cite{mea:84,desfrancois_prl:94} appear to be limited 
to the highest root $\gamma_{0,0}$ because of the characteristic order of 
magnitude of the molecular dipole moments realized in nature.

\section{Rotational Degrees of Freedom of Dipole-Bound Anions}
\label{sec:rotational_dipole-bound}

We now turn, through an appropriate length-scale hierarchy, 
to a derivation of the connection between the approach of
Refs.~\cite{bawin,Garrett_dip_rot,Clary_dip_rot,desfrancois_epj:98}
and the conformal treatment of Ref.~\cite{molecular_dipole_anomaly}.

\subsection{Rotationally Adiabatic Theory}
\label{sec:rotationally_adiabatic}

In the rotationally adiabatic theory~\cite{Clary_dip_rot},
the pseudopotential
\begin{equation}
\mathcal{V} (r)
=
- \frac{\hbar^{2}}{2 m_{e}}
\,  \frac{  
\Gamma 
\mbox{\boldmath\large  $\left(  \right.$ } \! \! \!
\lambda; F(r) 
\mbox{\boldmath\large  $\left.  \right)$ } \! \!
}{r^{2}}
\, G(r)
\;
\label{eq:effective_adiabatic_potential}
\end{equation}
for the radial electron wave function
is an eigenvalue of the reduced Hamiltonian
\begin{equation}
\hat{ \mathcal{H} }
=  - \frac{\hbar^{2}}{2 m_{e}}
  \, 
\frac{
\hat{ \mathcal{ A} } 
\mbox{\boldmath\large  $\left(  \right.$ } \! \! \!
\lambda; F (r) 
\mbox{\boldmath\large  $\left.  \right)$ } \! \!
}{r^{2}}
\, 
G(r)
\,
\; ,
\label{eq:adiabatic_Hamiltonian}
\end{equation}
and the radial function $G(r)$ can be selected by comparison 
with different expressions used in the 
literature~\cite{bawin,Garrett_dip_rot,Clary_dip_rot,desfrancois_epj:98}.
In particular, the lowest eigenvalue gives the standard 
adiabatic potential:
 $\epsilon_{\rm adiab} (r) \equiv 
\mathcal{V} (r)$.
In addition, the nontrivial part of the effective Hamiltonian of
Eq.~(\ref{eq:adiabatic_Hamiltonian}) arises from the
adiabatic approximation for the rotational motion of the molecule,
which provides the 
operator~\cite{bawin,Clary_dip_rot,desfrancois_epj:98}
\begin{equation}
\hat{ \mathcal{ A} } 
\mbox{\boldmath\large  $\left(  \right.$ } \! \! \!
\lambda; F (r) 
\mbox{\boldmath\large  $\left.  \right)$ } \! \!
=
- F(r) \,  {\bf l}^{2} 
+
\lambda\,  \cos \theta  
\; ,
\label{eq:adiabatic_angular_operator} 
\end{equation}
where the function $F(r) $ has the form  
$F(r) 
=
 1 + \left( r/r_{B} \right)^{2}
$, 
in which the length scale 
\begin{equation}
r_{B}  = \sqrt{ \frac{\hbar^{2} }{ 2 \, m_{e} \, B } }
\; 
\label{eq:IR_rotational_scale}
\end{equation}
is associated with the rotator constant $B = \hbar^{2}/2I$
(with $I$ being the moment of inertia).
Simple inspection shows that 
$
\hat{ \mathcal{ A} } 
\mbox{\boldmath\large  $\left(  \right.$ } \! \! \!
\lambda; F (r) 
\mbox{\boldmath\large  $\left.  \right)$ } \! \!
$
is a generalization of 
$\hat{ A } ( \lambda )$,
in which the replacement
${\bf l}^{2} \rightarrow F(r) \,  {\bf l}^{2} $ is made;
therefore, their angular operator structures are identical.
Using again the orbital angular momentum basis $\left| l,m \right\rangle$ 
of the electron,
the eigenvalue 
$\Gamma \equiv \Gamma(\lambda; F)$
of $\hat{ \mathcal{ A} } (  \lambda; F ) $ 
can be found from the secular equation
\begin{equation}
{\mathcal D}_{m} 
\mbox{\boldmath\large  $\left(  \right.$ } \! \! \!
\Gamma, \lambda; F(r) 
\mbox{\boldmath\large  $\left.  \right)$ } \! \!
\equiv
 \det 
\bigl[
\mathcal{M}_{ll'} 
\mbox{\boldmath\large  $\left(  \right.$ } \! \! \!
\Gamma, \lambda; m; F(r) 
\mbox{\boldmath\large  $\left.  \right)$ } \! \!
\bigr]
 = 0
\; ,
\label{eq:adiabatic_angular_determinant}
\end{equation}
where 
$\mathcal{M}  
\mbox{\boldmath\large  $\left(  \right.$ } \! \! \!
\Gamma, \lambda; F(r)  
\mbox{\boldmath\large  $\left.  \right)$ } \! \!
=
- 
{ \mathcal{ A} } 
\mbox{\boldmath\large  $\left(  \right.$ } \! \! \!
\lambda; F (r) 
\mbox{\boldmath\large  $\left.  \right)$ } \! \!
+ 
\Gamma   \, \openone$,
so that
$\mathcal{M}_{ll'} 
\mbox{\boldmath\large  $\left(  \right.$ } \! \! \!
\Gamma, \lambda; m; F(r) 
\mbox{\boldmath\large  $\left.  \right)$ } \! \!
$
is obtained from Eq.~(\ref{eq:conformal_tridiagonal_Mll'})
by the replacements $l(l+1) \rightarrow l(l+1) \, F(r)$
and $\gamma \rightarrow \Gamma$ in the diagonal terms.
Therefore the eigenvalues arising from 
Eq.~(\ref{eq:adiabatic_angular_determinant})
can be labeled just as those derived from the conformal
secular determinant:
$\Gamma_{h,m}$. 
In particular, the largest one, $\Gamma_{0,0}$, 
leads to the standard adiabatic potential
 $\epsilon_{\rm adiab} (r) =
- \hbar^{2}  \, 
\Gamma_{0,0} 
\mbox{\boldmath\large  $\left(  \right.$ } \! \! \!
\lambda; F(r) 
\mbox{\boldmath\large  $\left.  \right)$ } \! \!
G(r)/(2 m_{e} \, r^{2} )
$
in Eq.~(\ref{eq:effective_adiabatic_potential}).

\subsection{Separation of Scales: Renormalization Theory}
\label{sec:scales&renormalization}

The current reformulation of the rotationally adiabatic theory 
permits a direct comparison with the results of the 
conformal framework, to which it reduces by the use of 
effective field theory arguments. 
The reason for this lies in that,
in a renormalization treatment, 
the phenomenological factor $G(r)$ merely
amounts to an ultraviolet regulator---only 
needed for distances $r \alt a$, where $a$ is the size of 
the molecule. In other words,
the details of the position dependence of $G(r)$ are of secondary 
importance because $G(r) \approx 1$ for $r \agt a$
and the conformal potential effectively dominates the relevant physics.
Consequently, the only significant addition to the conformal framework 
appears to be the inclusion
of rotational degrees of freedom via the function $F(r)$.
However, a careful analysis of Eq.~(\ref{eq:adiabatic_angular_determinant}) 
shows that the conclusions from the conformal framework 
are not substantially altered.
The fundamental concept that underlies this surprising
result---and which makes our construction successful---is 
the clear-cut {\em separation of scales\/}. 
This is the essential assumption that underlies
renormalization theory~\cite{weinberg}, as described in the 
effective field theory language~\cite{polchinski}.
Specifically, the two characteristic length scales for the
molecular anions are
(i) a scale of the order of the molecular size $a$;
and (ii)
the rotational scale $r_{B}$
of Eq.~(\ref{eq:IR_rotational_scale}), whose size can be gleaned from
$I \sim M a^{2}$, with $M$ being the mass of the molecule.
Then, the scale hierarchy
\begin{equation}
r_{B} \sim \sqrt{ \frac{M}{m_{e} } } \; a 
\gg a
\; 
\label{eq:scales}
\end{equation}
shows that
$L_{\rm UV} \sim a$,
and 
$L_{\rm IR} \sim r_{B}$ 
play the role of ``ultraviolet'' and ``infrared'' scales
respectively. Moreover, Eq.~(\ref{eq:scales})
provides a justification for the adiabatic 
approximation used in
Refs.~\cite{bawin,Garrett_dip_rot,Clary_dip_rot};
remarkably, this approximation is just a statement about 
length scales within an effective-field-theory 
description of molecular physics~\cite{wilczek}.
Thus the conformal treatment constitutes
a satisfactory framework for the physics of dipole-bound molecular anions. 
This description can be further justified by introducing
a systematic reduction procedure.
First, the dependence 
of ${\mathcal V}(r)$ for $r \gg r_{B}$
plays a secondary role for the problem of criticality.
This can be rigorously established by an asymptotic analysis of the 
determinant~(\ref{eq:adiabatic_angular_determinant}).
Most importantly, 
the existence of a critical value and the ensuing bound
states follow from the relevant scales
$r \alt r_{B}$:
criticality does
{\em not\/} originate in the infrared sector.
Second, the critical dipole moment arises
from the ultraviolet boundary and can be established
by a renormalization framework.
Therefore the dominant physics
can be extracted by considering  
the intermediate scales,  with 
$
a \alt r  \ll  r_{B}$.
In that range,
 $F(r) \approx 1$ and $\Gamma (\lambda; F)$ 
in Eq.~(\ref{eq:adiabatic_angular_determinant})
can be replaced by a constant
$\gamma (\lambda) \equiv \Gamma (\lambda; 1)$.
Thus, in this ``scale window,''
the adiabatic potential approximately reduces to
a long-range conformal potential
${\mathcal V} (r)
=
- \hbar^{2} \gamma /(2 m_{e}r^{2})$. 
Retracing the previous steps,
this reduction establishes the 
Hamiltonian~(\ref{eq:ISP_Hamiltonian_unregularized_anisotropic}),
whose conformal symmetry is reminiscent of the corresponding 
description in high-energy physics~\cite{jac:72}:
at sufficiently small distances  
the problem becomes scale invariant.
Finally,
when a length scale of the order
$a$ is reached, ``new physics'' emerges and
a more detailed treatment is in order---for
which a specific form of the factor $G(r)$ would be needed.

\section{Generalized Conformal Framework:
Predictions and Nature of the Corrections}
\label{sec:generalized_renormalization}

The length-scale analysis
leads to a noteworthy adjustment to the previous results:
the restriction of the conformal 
tower of bound states to the relevant range of scales.
This is due to the fact that
the dominant physics is described by a ``conformal
window'' limited by the characteristic scales
$L_{\rm UV}$
and
$L_{\rm IR}$, which
act as ultraviolet and infrared cutoffs~\cite{camblong_CQM}.
The existence of an 
ultraviolet boundary is directly involved in the renormalization process
and drives the fundamental properties
of the renormalized conformal framework.
By contrast, as shown in Ref.~\cite{camblong_CQM},
the infrared boundary 
only restricts the range of the dominant physics.

Most importantly, there are a number of
predictions arising from this {\em generalized conformal framework\/},
which---with appropriate refinements---could be tested experimentally
and compared against results from alternative approaches. 
We will illustrate these results
by considering the dominant sector of the theory in the subspace
${\mathcal S}_{m=0}(l=0,1)$
of quantum  numbers $l=0$ and $l=1$ for the secular  
determinant~(\ref{eq:adiabatic_angular_determinant}) with $m=0$,
in which
$\Gamma_{0,0}
=
- F 
+
\sqrt{
F^{2} + \lambda^{2}/3
}$~\cite{Clary_dip_rot}.

The first prediction arises directly from
 the existence of a conformal domain, which implies that
the number of {\em conformal bound states\/} undergoes
a cutoff process leading to a finite value $N_{\rm conf} $.
It turns out that the approximate number
\begin{equation}
N_{\rm conf} 
\sim \frac{\Theta}{\pi} 
\,
\ln \left( \frac{L_{\rm IR}}{L_{UV} } \right) 
\;  ,
\label{eq:number_conformal_states}
\end{equation}
which is predicted from renormalization,
is also in good agreement with
known bound-state estimates~\cite{Calogero,number_of_states}.
For typical values of the parameters involved,
the logarithmic nature of $N_{\rm conf} $ yields 
the generally accepted result 
that dipole-bound molecular anions sustain 
only one or two bound states. 
Therefore, in contrast with the claims of Ref.~\cite{bawin},
our approach shows that the presence of a conformal domain
is the actual {\em cause for the existence of bound states
and of the critical dipole moment\/}.

The second important prediction of the generalized
renormalization framework consists of
corrections to the critical value  $\lambda^{(*)}$.
Within the effective-field reduction, as a zeroth-order approximation,
Eq.~(\ref{eq:adiabatic_angular_determinant})
[with  $F(r) \approx 1$]
provides the required critical dimensionless dipole moment 
$\lambda^{(*)}_{\rm conf}$,
which is purely conformal in nature.
Broadly speaking, when a dipole moment is 
sufficiently different from the critical value,
the predictions of the conformal framework are remarkably accurate.
However, very near criticality, $\Theta \sim 0 $ and 
$\kappa \sim 0$; 
this is due to the fact that
the condition of criticality amounts to the 
{\em emergence of a ground state from the continuum\/}.
The corresponding enlarged characteristic size of the ground-state
conformal wave function
links the relevant scales and corrections are unavoidable
in the presence of an infrared cutoff.
One possible way of dealing with this is through a perturbative
evaluation of $\lambda^{(*)}$ at the level of
Eq.~(\ref{eq:adiabatic_angular_determinant});
nevertheless, because of the
extremely long range of the wave function~(\ref{eq:radial_wave_function}),
one would have to consider all orders of perturbation theory 
and carry out infinite resummations.
An alternative,
 more direct estimate can be established 
from the emergence of the first bound state,
\begin{equation}
N =
N_{\rm conf} 
+
\delta
= 
1
\; ,
\label{eq:emergence_GS}
\end{equation}
where 
$
\delta 
= 
\delta_{\rm IR} + \delta_{\rm UV} 
$
is the partial contribution of the infrared and ultraviolet 
sectors to the number of states.
The criticality condition~(\ref{eq:emergence_GS}), combined with
Eq.~(\ref{eq:number_conformal_states}),
can then be used to evaluate the conformal parameter
$\Theta_{\rm gs}$ of the critical
ground-state wave function; the fact that $\Theta_{\rm gs}$ 
is small but finite is due to the self-consistent 
restriction of the theory in the infrared. 
Thus the fractional correction to the critical dipole value
\begin{equation}
\epsilon 
\equiv
\frac{\lambda^{(*)}}{
\lambda^{(*)}_{\rm conf} } - 1
\; 
\label{eq:modified_critical_dipole_def}
\end{equation}
can be computed from the secular 
equation~(\ref{eq:adiabatic_angular_determinant}),
by means of Eq.~(\ref{eq:Theta_definition}),
in which $\gamma = 1/4$ for the purely conformal theory,
while 
$\tilde{\gamma} 
= 
\Theta_{\rm gs}^{2} +
1/4
$ for the theory with an infrared cutoff,
so that
\begin{equation}
\Delta \gamma = \tilde{\gamma } -  \frac{1}{4} =
\Theta_{\rm gs}^{2} =
4 \pi^{2} 
\left( 1 - \delta \right)^{2}
\,
\left[ \ln \left( \frac{ r_{B} }{ a } \right)^{2} 
\right]^{-2}
\;  .
\label{eq:modified_critical_dipole_basic}
\end{equation}
In particular, in the
restriction of the theory to the dominant subspace
${\mathcal S}_{m=0}(l=0,1)$, the quantity
$\epsilon$ in Eq.~(\ref{eq:modified_critical_dipole_def}) becomes
\begin{equation}
\epsilon 
=
\sqrt{ 
[
1 + 4 (\tilde{\gamma} -\gamma )
]
\,
[
1 + \frac{4}{9} (\tilde{\gamma} -\gamma )
]
\,
}
\,
- 1
\approx \frac{20}{9} 
(\tilde{\gamma} -\gamma )
\; ,
\label{eq:restricted_subspace_epsilon}
\end{equation}
where the approximate equality
arises from the relatively small values
of
$
(\tilde{\gamma} -\gamma )$, which are
due to the {\em separation of scales\/}.
Consequently,
Eqs.~(\ref{eq:modified_critical_dipole_basic})
and (\ref{eq:restricted_subspace_epsilon}) imply that
\begin{equation}
\epsilon 
\approx 
\frac{20}{9} \, \Theta_{\rm gs}^{2} 
\approx
\frac{80 \,  \pi^{2}}{9} \,
\left( 1 - \delta \right)^{2}
\,
\left[ \ln \left( \frac{ r_{B} }{ a } \right)^{2} 
\right]^{-2}
\;  .
\label{eq:modified_critical_dipole}
\end{equation}
As expected,
this correction becomes more prominent for
decreasing values of $I$
and increases the critical dipole from its ideal
conformal value.
In addition, the fractional state contribution $\delta$
in the compensatory factor $(1-\delta)$
can be determined using standard estimates for the 
number of bound states~\cite{number_of_states}.
With these building blocks, Eq.~(\ref{eq:modified_critical_dipole}) 
gives the leading dependence of the critical value $\lambda^{(*)}$
with respect to the infrared scale
through
$\ln \rho$, 
with $\rho \equiv I/(m_{e} a^{2}) = r_{B}^{2}/a^{2}$
being the dimensionless
molecular moment of inertia. The logarithmic dependence
$\ln \rho$ is
the {\em trademark of the underlying renormalization-induced physics\/}
and explains the slow convergence of 
$\lambda^{(*)}$ towards 
$
\lambda^{(*)}_{\rm conf}  
$. 
This analysis ultimately shows that,
even when rotational degrees of freedom 
are included in the description of this problem,
renormalization is still responsible for the predicted values of
$p^{(*)}$, including:

(i) the existence of a critical value
whose order of magnitude is given by
the conformal critical point~(\ref{eq:conformal_critical_point});
and

(ii) 
the underlying physics of the
logarithmic correction~(\ref{eq:modified_critical_dipole}).
\\
Most importantly, the 
results~(\ref{eq:number_conformal_states})-(\ref{eq:modified_critical_dipole})
are {\em universal, i.e., model-independent,
within the conformal framework\/}.

In addition, we acknowledge the existence of model-dependent corrections to
this framework. For molecular dipole anions, these effects can be represented
by means of a pseudopotential comprised of electrostatic 
terms---described by the multipole expansion---combined 
with many-body contributions of two kinds:
a polarization part and an exchange part due to the Pauli exclusion 
principle~\cite{Garrett_pseudopot,desfrancois_epj:98,desfrancois_prl:94,desfrancois_ijmpb:96,clusters}.
The long-distance electrostatic and polarization terms 
do not substantially affect the rotational infrared 
corrections to the purely conformal problem
because their coupling constants are proportional 
to $a^{2}$ (with the relevant rotational degrees of freedom being
proportional to $r_{B}^{2}$, and 
$r_{B} \gg a$)~\cite{desfrancois_epj:98,desfrancois_ijmpb:96}.
The short-distance behavior, which contributes to the ultraviolet physics 
with a scale of the order of $L_{\rm UV} \sim a$, 
involves electrostatic and exchange many-body 
effects~\cite{desfrancois_epj:98,desfrancois_ijmpb:96}.
In the case of the exchange effects, the characteristic scale is 
determined by the overlap of orbitals associated with tightly bound electrons,
and the corresponding repulsive core is highly dependent on the 
nature of the molecular species~\cite{exchange}, 
with $\delta_{\rm UV}<0$.
This negative value partially compensates the positive term
$\delta_{\rm IR}$ and favors the agreement with the observed critical dipole 
moment in complex molecular species. 
Consequently, the scale analysis confirms the 
remarkable fact that {\em the dipole-bound anionic state exists primarily 
due to the conformal interaction\/}~\cite{SGS:99}.
One of the simplest characterizations of these model-dependent corrections
is afforded by the dominant limiting infrared
behavior of the rotationally adiabatic theory of 
Ref.~\cite{Clary_dip_rot}, which yields
$\delta \approx \delta_{\rm IR}
\approx
\sqrt{6} \, \lambda^{(*)}_{\rm conf}  
\,
\left(
1 + \epsilon
\right)/3 \pi
$.
With these assignments,
introducing the parameters $c 
=
\left[
\left( \sqrt{6} \, \lambda^{(*)}_{\rm conf}/3 \pi \right)^{-1}
- 1 \right]^{-1}
\approx 
0.498$,
$A= 
80 \pi^{2}L^{-2}/[9(c+1)^{2}]$,
and $L
= \ln \rho
$,
the fractional correction to the 
dipole moment 
becomes
$\epsilon
\approx \left\{
[1+1/(2cA)]
- \sqrt{ [1+1/(2cA)]^{2} -1}
\right\}/c$;
for example,
for various values of the 
 dimensionless
molecular moment of inertia:
 $\rho = 2 \times 10^{8}$,
$\rho = 2 \times 10^{6}$,
and
$\rho = 4 \times 10^{4}$,
the corresponding fractional corrections are, respectively,
$\epsilon
\approx 0.11$,
$\epsilon \approx 0.16$, and $\epsilon \approx 0.26$~\cite{epsilon_garrett}.

Finally, let us consider another universal prediction
for an experimental realization with 
at least two conformal
bound states~\cite{third_state}.
For such a system, 
Eq.~(\ref{eq:ratios_cutoff_BS_regularized_energies_phenomenological})
yields the ratio
$E_{1}/E_{0} = \exp \left( - 2 \pi/\Theta \right) $
from which the relative value
of the dipole moment, compared to the critical dipole, is
\begin{equation}
\frac{\lambda}{
\lambda^{(*)} } - 1
\approx \frac{20}{9} \, \Theta^{2} = 
\frac{80 \,  \pi^{2}}{9} \,
\left[ \ln \left( \frac{E_{1}}{E_{0}} \right) \right]^{-2}
\;  ,
\label{eq:critical_dipole_inverted}
\end{equation}
which can be derived with the restriction to  ${\mathcal S}_{m=0}(l=0,1)$,
and supplemented by critical-dipole corrections
just as in Eq.~(\ref{eq:modified_critical_dipole}).
This ``inversion'' makes a simple prediction 
solely based on conformal quantum mechanics
and which can be explicitly compared against the 
improved critical value~(\ref{eq:modified_critical_dipole}), using
the known dipole moment $\lambda$ for the given
polar molecule.
In essence, this is a test of the residual scale invariance of the geometric 
scaling~(\ref{eq:ratios_cutoff_BS_regularized_energies_phenomenological})
of the conformal tower of states.

\section{Conclusions}
\label{sec:conclusions}

In conclusion, the central concept put forward in this paper
is the 
{\em anomalous
emergence of bound states via renormalization\/}
for a system with a conformally invariant domain
whose ultraviolet boundary dictates binding.
The ensuing quantum symmetry breaking within this framework
captures the essence of the observed critical 
dipole moment for the formation of dipole-bound anions.

Moreover,
the tools developed in this paper,
as exemplified by
Eqs.~(\ref{eq:number_conformal_states})-(\ref{eq:critical_dipole_inverted}),
show that this conformal framework:

(1) 
 permits the extraction
of {\em universal properties\/} 
for physical problems 
with a conformally invariant domain;
and

(2)
provides a
description of dipole-bound anions in which model-dependent
and model-independent contributions can be conveniently organized.

In principle,
this generalized
conformal framework could be used as the starting point of
a systematic approximation scheme for the description of
dipole-bound molecular anions.
The estimate~(\ref{eq:modified_critical_dipole})
is a typical illustration of this: its
 numerical coefficients
could be further refined by an improved {\em matching \/}
of the conformal domain with the infrared and ultraviolet
sectors, as well as by considering higher orders
(with respect to $l$).
Thus our problem is similar to that encountered in many other
areas of physics, in which
a zeroth order approximation
captures the essential ingredients,
which are to be subsequently improved upon by the use of 
miscellaneous approximation techniques.

Most intriguingly, 
our approach exhibits many similarities 
with the recently developed chiral-Lagrangian program for
nuclear physics~\cite{weinberg_nuclear,ordonez},
in which the underlying chiral
symmetry from QCD provides a guiding principle within
a power-counting scheme that selects the 
terms in the Lagrangian for nucleons and pions---with
the first terms 
capturing the dominant, model-independent
contributions.
Likewise, our conformal framework,
based on the SO(2,1) invariance
and the use of effective-field
theory concepts,
 is a discriminating scheme
to elucidate the dominant model-independent
features of the molecular anions and similar
 systems with a conformally invariant domain;
in this context, it would be interesting to
develop the analog of the chiral
 power-counting scheme.

\acknowledgments{This work was supported by CONICET, ANPCyT, 
the University of San Francisco Faculty Development Fund,
and by the National Science Foundation under Grants
No.\ 0308300
and
No.\ 0308435.}


\begin{thebibliography}{99}
\bibitem{number_of_states}
In this paper we are using
the upper bound 
$(2/\pi) \,
\int dr \, | 2 m_{e} {\mathcal V}(r)/\hbar^{2}|^{1/2}$;
see Ref.~\cite{Calogero}.
\bibitem{epsilon_garrett}
The corresponding numerical values in the first article
in Ref.~\cite{Garrett_dip_rot}
(Table I, p. 967) 
are 0.11, 0.16, and 0.23, respectively
(for $R= 1.0284 \, a_{0}$ and $J=0$).
\end{thebibliography}
\end{document}